\documentclass[a4paper,10pt]{article}
\usepackage[latin1]{inputenc}
\usepackage{amssymb}
\title{{\bf  On infinite real trace rational  languages  
of maximum topological complexity} \thanks{The results of this paper have been exposed 
during  the International Conference JAF 22,  22nd Journées sur les Arithmétiques Faibles, 
June 11-14, 2003, Napoli, Italy. 
}}
\author{$^1$Olivier Finkel and  $^2$Jean-Pierre Ressayre and 
$^3$Pierre Simonnet\\
\\
{\it $^{1,2}$Equipe de Logique Math\'ematique }  
 \\ U.F.R. de Math\'ematiques, Universit\'e Paris 7 \\ {\it 2 Place Jussieu 75251 Paris
 cedex 05, France.}
\\ finkel@logique.jussieu.fr ~~ressayre@logique.jussieu.fr~~
\\ {\it $^{3}$UMR CNRS 6134,} Facult\'e des Sciences, Universit\'e de Corse\\
{\it Quartier Grossetti BP52 20250, Corte, France } 
\\ 
simonnet@univ-corse.fr }
\date{}
\begin{document}

\newtheorem{The}{Theorem}[section]
\newtheorem{Pro}[The]{Proposition}
\newtheorem{Deff}[The]{Definition}
\newtheorem{Lem}[The]{Lemma}
\newtheorem{Rem}[The]{Remark}
\newtheorem{Exa}[The]{Example}

\newcommand{\fa}{\forall}
\newcommand{\Ga}{\Gamma}
\newcommand{\Gas}{\Gamma^\star}
\newcommand{\Gao}{\Gamma^\omega}

\newcommand{\Si}{\Sigma}
\newcommand{\Sis}{\Sigma^\star}
\newcommand{\Sio}{\Sigma^\omega}
\newcommand{\ra}{\rightarrow}
\newcommand{\hs}{\hspace{12mm}

\noi}
\newcommand{\lra}{\leftrightarrow}
\newcommand{\la}{language}
\newcommand{\ite}{\item}
\newcommand{\Lp}{L(\varphi)}
\newcommand{\abs}{\{a, b\}^\star}
\newcommand{\abcs}{\{a, b, c \}^\star}
\newcommand{\ol}{ $\omega$-language}
\newcommand{\orl}{ $\omega$-regular language}
\newcommand{\om}{\omega}
\newcommand{\nl}{\newline}
\newcommand{\noi}{\noindent}
\newcommand{\tla}{\twoheadleftarrow}
\newcommand{\de}{deterministic }
\newcommand{\proo}{\noi {\bf Proof.} }
\newcommand {\ep}{\hfill $\square$}
\renewcommand{\thefootnote}{\star{footnote}}

\maketitle 
\begin{abstract}
\noi We consider the set $\mathbb{R}^\om(\Ga, D)$ of infinite real traces, 
over a dependence alphabet $(\Ga, D)$ with no isolated letter, 
equipped with the topology induced by the prefix metric.  
We then prove  that all rational languages of infinite real traces are analytic sets. 
We reprove also that there exist some  rational  languages of  infinite real traces
which are analytic but non Borel sets, and even ${\bf \Si^1_1}$-complete, 
hence of maximum possible topological complexity. For that purpose we give an example 
of ${\bf \Si^1_1}$-complete language which is fundamentally different from the known example 
of ${\bf \Si^1_1}$-complete infinitary rational relation given in \cite{relrat}. 
\end{abstract}

\noi {\small {\bf Keywords:} Real traces;  rational languages; topological properties; analytic 
and Borel sets.}

\section{Introduction}

Trace monoids were firstly considered by Cartier and Foata for studying combinatorial problems, 
\cite{cf}. Next Mazurkiewicz introduced finite traces as a semantic model for 
concurrent systems, \cite{maz}. Since then traces have been much investigated by 
various authors and they have been extended to 
infinite traces  to model systems which may not terminate;  see the handbook \cite{handbook} and 
its chapter about infinite traces \cite{gp}, for many results and references. 
\nl In particular,  
{\bf real traces}  have been studied by Gastin,  Petit and  Zielonka, 
who characterized in  \cite{gpz}  
the two important families of recognizable and rational languages of  real traces, over 
a dependence alphabet $(\Ga, D)$, in connection with rational languages of 
finite or infinite words. 
\nl  Several metrics have been defined on the set  $\mathbb{R}(\Ga, D)$ of real traces 
over $(\Ga, D)$. In particular,   the prefix metric defined by Kwiatkowska \cite{kw} and the 
 Foata normal form metric defined by Bonnizzoni,  Mauri and Pighizzini  \cite{bmp}.
 Kummetz and  Kuske stated in \cite{kk} that for 
{\it finite} dependence alphabets these two metrics 
define the same topology on $\mathbb{R}(\Ga, D)$. Moreover, if we consider only 
{\bf infinite real traces} over a dependence alphabet $(\Ga, D)$ without isolated letter, 
the topological subspace 
$\mathbb{R}^\om(\Ga, D)=\mathbb{R}(\Ga, D)-\mathbb{M}(\Ga, D)$ of $\mathbb{R}(\Ga, D)$ 
(where $\mathbb{M}(\Ga, D)$ 
is the set of finite traces over $(\Ga, D)$), is homeomorphic to the Cantor set 
$2^\om$, or equivalently to any  set  $\Sio$ of infinite words 
over a finite alphabet $\Si$, equipped with the product of the discrete topology on  $\Si$, 
\cite{kk,sta,pp}.  
\nl  We can then define, from open subsets of
 the topological space $\mathbb{R}^\om(\Ga, D)$, the hierarchy of Borel 
sets by successive operations of countable intersections and countable unions. Furthermore, 
it is well known that there exist some subsets of the Cantor set, 
hence also some subsets of  $\mathbb{R}^\om(\Ga, D)$, which are not Borel. There is another 
hierarchy beyond the Borel one, called the projective hierarchy. 

\hs It is then natural to try to locate classical languages of infinite real traces 
with regard to these hierarchies and this question is posed by Lescow and Thomas 
in \cite{lt} (for the general case 
of infinite labelled partial orders like traces). 
 In the case of infinite words,  Mc Naughton's Theorem 
implies that every \orl~ is a 
boolean combination of ${\bf \Pi^0_2 }$-sets 
hence a ${\bf \Delta^0_3 }=({\bf \Pi^0_3 } \cap {\bf \Pi^0_3 })$-set.  Landweber studied first 
the topological properties of \orl s and 
 characterized the \orl s in each of the Borel classes 
 ${\bf \Si^0_1 }$, ${\bf \Pi^0_1 }$,  ${\bf \Si^0_2 }$,   ${\bf \Pi^0_2 }$,  \cite{la}.
We study in this paper the topological complexity of {\bf rational languages of 
infinite real traces}.  
\nl  We   show below that all rational languages of infinite real traces 
are analytic sets and that
 there exist some  rational  languages of  infinite real traces
which are analytic but non Borel sets, and even ${\bf \Si^1_1}$-complete, 
hence of maximum possible topological complexity, giving a partial answer to the question 
of the comparison between the
 topological complexity of rational languages of infinite words and of 
infinite traces \cite{lt}. 

\hs The first author recently showed in \cite{relrat} that there exists a 
${\bf \Si^1_1}$-complete infinitary rational relation 
$R \subseteq \Si_1^\om \times \Si_2^\om$ where $\Si_1$ and $\Si_2$ are two finite alphabets 
having at least two letters. 
\nl We could have used this result to prove that there exists a ${\bf \Si^1_1}$-complete 
rational language of infinite real traces $L \subseteq \mathbb{R}^\om(\Ga, D)$, whenever 
$(\Ga, D)$ is a dependence alphabet and $\Ga \supseteq \Si_1 \cup \Si_2$,  where 
$\Si_1$ and $\Si_2$ are two independent dependence cliques having at least two letters. 
This can be done by considering the 
natural embedding $i: \Si_1^\om \times \Si_2^\om \ra \mathbb{R}^\om(\Ga, D)$. The language 
$R'=i(R)$ is then ${\bf \Si^1_1}$-complete. 
 But this way the language  $R'$ would have in fact the  structure of an 
infinitary rational relation. 
\nl On the other side the  
${\bf \Si^1_1}$-complete language $\mathcal{L}$ given in this paper is a new example 
whose structure is radically  different from that of $R'$. 
In particular,  $\mathcal{L}$ does not contain any ${\bf \Si^1_1}$-complete language of infinite 
traces having the structure of an infinitary rational relation. 
\nl This is important because in Trace Theory the structure of dependence alphabets is very 
important: some results are known to be true for some dependence alphabets and false 
for other dependence alphabets (see for example \cite{hm}). 

\hs  Moreover we think that the presentation of this new example 
has also some interest for the following reasons. 
 The proof given in this paper is self contained and is stated in the general context 
of traces. The problem is exposed in this general context and  we use here the general 
properties of traces instead of the particular properties of infinitary rational relations. 
We use the characterization of rational languages of 
infinite real traces given by Theorem \ref{rat} of Gastin, Petit and Zielonka, which 
states a connection between rational languages of infinite words and rational languages 
of infinite real traces instead of the notion of B\"uchi transducer. 

\hs  We prove also that all rational languages of infinite real traces are analytic. 
Our proof is not difficult, but it is original, using the Baire space $\om^\om$ and the 
characterization of rational languages of 
infinite real traces given by Theorem \ref{rat}: every rational  
language of infinite traces is  a finite union of sets of the 
form $R.S^\om$ where $S$ and $R$ are rational 
monoalphabetic languages of finite traces. 
This proof is rather different from 
usual ones in the theory of $\om$-languages. 
\nl It uses a connection between  the topological complexity of $\om$-powers of  
languages of finite  traces, i.e. of languages of the form $S^\om$, where $S$ is a 
language of finite traces, and the topological complexity of rational 
languages of infinite traces. The closure under countable union of the class 
of analytic sets is also important in the proof. 
\nl We expect that the inverse way could also be fruitful: it seems to us that 
there should exist some context free $\om$-languages and some infinitary rational relations 
of high transfinite Borel rank. We think that we could 
use this fact to show that there exist $\om$-powers of finitary languages 
of high transfinite Borel rank. 
Notice that the question of the topological complexity of $\om$-powers (of languages of finite 
words) has been raised by 
several authors \cite{Niwinski90,Simonnet92,sta,Staiger97b} 
and some new results have been recently proved 
\cite{Fin01a,finb,op,Lecomte01}.

\hs 
The paper is organized as follows. In section 2 we recall the notion of words and traces. 
In section 3 we recall definitions of Borel 
and analytic sets, and we prove our main results in section 4.

\section{Words and traces}

\noi Let us now introduce notations for words.
 For  $\Si$  a finite alphabet,   
a non empty finite word over $\Si$ is a finite sequence of letters:
 $x=a_1a_2\ldots a_n$ where $\fa i\in [1; n]$ $a_i \in\Si$.
 We shall denote $x(i)=a_i$ the $i^{th}$ letter of $x$
and $x[i]=x(1)\ldots x(i)$ for $i\leq n$. The length of $x$ is $|x|=n$.
The empty word will be denoted by $\varepsilon$ and has no letters. Its length is 0.
 The set of non empty finite words over $\Si$ is denoted $\Si^+$.
 $\Sis = \Si^+  \cup \{\varepsilon\}$ is the set of finite words over $\Si$.
 A (finitary) language $L$ over $\Si$ is a subset of $\Sis$.
 The usual concatenation product of $u$ and $v$ will be denoted by $u.v$ or just  $uv$.
 For $V\subseteq \Sis$, we denote  \quad 
$V^\star=\{ v_1\ldots v_n  \mid  
n\geq 1 \quad and \quad \fa i \in [1; n] \quad v_i \in V  \}\cup \{\varepsilon\}$.

\hs The first infinite ordinal is $\om$.
An $\om$-word over $\Si$ is an $\om$ -sequence $a_1a_2\ldots a_n \ldots$, where 
$\fa i\geq 1 ~a_i \in\Sigma$.
 When $\sigma$ is an $\om$-word over $\Si$, we write
 $\sigma =\sigma(1)\sigma(2)\ldots  \sigma(n) \ldots $
and $\sigma[n]=\sigma(1)\sigma(2)\ldots  \sigma(n)$ the finite word of length $n$, 
prefix of $\sigma$.
The set of $\om$-words over  the alphabet $\Si$ is denoted by $\Si^\om$.
 An  $\om$-language over an alphabet $\Sigma$ is a subset of  $\Si^\om$.
For $V\subseteq \Sis$,
 $V^\om = \{ \sigma =u_1\ldots  u_n\ldots  \in \Si^\om \mid  \fa i\geq 1 ~~u_i\in V \}$
is the $\om$-power of $V$.
 The concatenation product is extended to the product of a 
finite word $u$ and an $\om$-word $v$: 
the infinite word $u.v$ is then the $\om$-word such that:
 $(u.v)(k)=u(k)$  if $k\leq |u|$ , and  $(u.v)(k)=v(k-|u|)$  if $k>|u|$.
\nl The prefix relation is denoted $\sqsubseteq$: the finite word $u$ is a prefix of the finite 
word $v$ (respectively,  the infinite word $v$), denoted $u\sqsubseteq v$,  
 if and only if there exists a finite word $w$ 
(respectively,  an infinite word $w$), such that $v=u.w$.
\nl We shall denote $\Si^\infty = \Sis \cup \Sio$ the set of finite or infinite words over 
$\Si$. 

\hs We introduce firstly traces as dependence graphs, \cite{handbook,gp,kk}. 
A dependence relation over 
an alphabet $\Ga$ is a reflexive and symmetric relation on $\Ga$.  Its complement 
$I_D=(\Ga \times \Ga)-D$ is the independence relation induced by the relation $D$; the relation 
$I_D$ is  irreflexive and symmetric.  
A dependence alphabet  $(\Ga, D)$ is formed by a finite alphabet $\Ga$ and a 
dependence relation $D \subseteq \Ga \times \Ga$. 
\nl A dependence graph $[V, E, \lambda]$ over the dependence alphabet $(\Ga, D)$ is an 
isomorphism class of a node labelled graph $(V, E, \lambda)$ such that $(V, E)$ is a 
directed acyclic graph, $V$ is at most countably infinite, $\lambda: V  \ra \Ga$ is a function 
which associates 
a label $\lambda(a)$ to each node $a\in V$, and such that:
\begin{enumerate}
\ite[(1)] $\fa v, w \in V ~~~(\lambda(v), \lambda(w)) \in D \leftrightarrow 
(v=w  \mbox{ or } (v, w) \in E  \mbox{ or } (w, v) \in E)$ 
\ite[(2)] The reflexive and transitive closure $E^\star$ of the edge relation 
$E$ is well founded, i.e. there is no infinite strictly decreasing sequence of vertices.  

\end{enumerate} 

\noi Let us remark that, since in this definition  $(V, E)$ is acyclic,   
 $E^\star$ is a partial 
order on $V$. 

\hs The empty trace has no vertice and will be denoted by $\varepsilon$ as in the case of words.

\hs As usually  the concatenation of two dependence graphs $g_1=[V_1, E_1, \lambda_1]$ and 
$g_2=[V_2, E_2, \lambda_2]$, where we can assume,  without loss of generality,  that 
$V_1$ and $V_2$ are disjoint, is the dependence graph $g_1.g_2=[V, E, \lambda]$ such that 
$V=V_1\cup V_2$, $E=E_1 \cup E_2 \cup \{(v_1, v_2)\in V_1\times V_2 \mid 
(\lambda_1(v_1), \lambda_2(v_2)) \in D\}$, and $\lambda=\lambda_1 \cup \lambda_2$. 

\hs The alphabet $alph(t)$ of a trace $t=[V, E, \lambda]$ is the set $\lambda(V)$. The 
alphabet at infinity of $t$ is the set 
$alphinf(t)=\{a\in \Ga \mid \lambda^{-1}[a] \mbox{ is infinite }\}$ of all $a\in alph(t)$ 
occurring infinitely often in $t$. 

\hs The set $\mathbb{M}(\Ga, D)$ of finite traces over  $(\Ga, D)$ is the set of traces having 
only finitely many vertices. 
For $t\in \mathbb{M}(\Ga, D)$ the length of $t$ is the number of vertices of $t$ denoted 
$|t|$.  
\nl The star operation $T \ra T^\star$ and the operation $T \ra T^\om$ are naturally 
extended to subsets $T$ of $\mathbb{M}(\Ga, D)$:
\nl $T^\star = \{ t_1.t_2 \ldots t_n  \mid  
n\geq 1 \mbox{ and }  \fa i \in [1; n] ~~ t_i \in T  \}\cup \{\varepsilon\}$
\nl $T^\om =  \{ t_1.t_2 \ldots t_n  \ldots \mid  \fa i~~ t_i \in T \}$.

\hs A real trace over $(\Ga, D)$ is a dependence graph $[V, E, \lambda]$ such that for all 
$v\in V$ the set $\{u\in V \mid (u, v) \in E^\star \}$ is finite. The set of real traces 
over $(\Ga, D)$ is denoted $\mathbb{R}(\Ga, D)$ and  the set 
$\mathbb{R}(\Ga, D)-\mathbb{M}(\Ga, D)$ 
of {\bf infinite} real traces
will be  denoted by  $\mathbb{R}^\om(\Ga, D)$.

\hs The prefix order over words can be extended to real traces in the following way. 
For all $s, t \in \mathbb{R}(\Ga, D)$  $s \sqsubseteq t$ iff there exists 
$z\in \mathbb{R}(\Ga, D)$ such that $s.z = t$ iff $s$ is a downwards closed subgraph of $t$.  
The corresponding suffix $z$ is then unique. 

\hs  Real traces may also be viewed as equivalence classes of (finite or infinite) words. 
Let $(\Ga, D)$ be a  dependence alphabet and let $\varphi: \Ga^\infty \ra \mathbb{R}(\Ga, D)$ 
be the mapping defined by $\varphi(a)=[\{x\}, \emptyset, x\ra a]$ for each $a\in \Ga$ and 
$\varphi(a_1.a_2\ldots )=\varphi(a_1).\varphi(a_2)\ldots $ for each $a_1.a_2\ldots $ in 
$\Ga^\infty$. Let us remark that if there is some $(a, b)$ in $I_D$ then the mapping is not 
injective because for instance $\varphi(ab)= \varphi(ba)$. One can define an equivalence 
relation $\sim_I$ on $\Ga^\infty$ by: for all $u, v \in \Ga^\infty$~~ $u \sim_I v$ iff 
$\varphi(u) = \varphi(v)$.  
Then $\varphi$ induces  a surjective morphism from the free monoid $\Gas$ onto the 
monoid of finite dependence graphs $\mathbb{M}(\Ga, D)=\Gas /\sim$. 
And the set $\varphi(\Ga^\infty)= \Ga^\infty  /\sim$ is the set of  real traces 
$\mathbb{R}(\Ga, D)$. 
\nl The empty trace is the image $\varphi(\varepsilon)$ of the empty word and is still 
denoted by $\varepsilon$.

\hs We assume the reader to be familiar with the theory of formal languages and of 
\orl s, see  \cite{tho,sta,pp} for many results and references. 
We recall that \orl s are accepted 
by B\"uchi  automata and that the class of \orl s  is the omega Kleene 
closure of the class of regular finitary languages. 
\nl The family of rational real trace languages over $(\Ga, D)$ is the smallest family 
which contains the emptyset, all the singletons 
$\{[\{x\}, \emptyset, x\ra a]\}$,  
 for $a\in \Ga$, and which is closed 
under finite union, concatenation product, $\star$-iteration 
and $\om$-iteration on real traces. 
\nl Let us now recall the following characterization of rational languages of 
infinite real traces (there exists also a version for finite {\it or}
 infinite traces, \cite{gpz}). A real trace language $R$ is said to be monoalphabetic 
if $alph(s)=alph(t)$ for all $s, t\in R$. 

\begin{The}[\cite{gpz}]\label{rat}
Let $T \subseteq \mathbb{R}^\om(\Ga, D)$ be a language of infinite real traces over 
the dependence alphabet  $(\Ga, D)$. The following assertions are equivalent:
\begin{enumerate}
\ite[(1)] T is rational.  
\ite[(2)] T is a finite union of sets of the form $R.S^\om$ where $S$ and $R$ are rational 
monoalphabetic languages of finite traces over $(\Ga, D)$ and $\varepsilon \notin S$. 
\ite[(3)] $T=\varphi(L)$ for some \orl~ $L \subseteq \Gao$.

\end{enumerate}

\end{The}

\section{Topology}

\noi We assume the reader to be familiar with basic notions of topology which
may be found in  \cite{kec,lt,sta,pp}.
\nl There is a natural metric on the set $\Sio$ of  infinite words 
over a finite alphabet 
$\Si$ which is called the prefix metric and defined as follows. For $u, v \in \Sio$ and 
$u\neq v$ let $d(u, v)=2^{-l_{pref(u,v)}}$ where $l_{pref(u,v)}$ is the first integer $n$
such that the $(n+1)^{th}$ letter of $u$ is different from the $(n+1)^{th}$ letter of $v$. 
This metric induces on $\Sio$ the usual  Cantor topology for which open subsets of 
$\Sio$ are in the form $W.\Si^\om$, where $W\subseteq \Sis$.
\nl (Notice that this prefix metric may be extended to the set $\Si^\infty$ 
of finite or infinite words over the alphabet $\Si$). 

\hs The prefix metric has been extended to real traces by Kwiatkowska in 
\cite{kw} by defining firstly for all $s, t \in  \mathbb{R}(\Ga, D)$ with $s\neq t$:
$$l_{pref}(s, t)=sup\{n\in \mathbb{N} \mid r \sqsubseteq s \lra r \sqsubseteq t \mbox{ for all }
r\in \mathbb{M}(\Ga, D) \mbox{ with } |r| \leq n \}$$

\noi and next $d_{pref}(s,t)=2^{-l_{pref(s,t)}}$

\hs Notice that we consider in this paper {\it infinite} traces and Kwiatkowska
defined the prefix metric over {\it finite or infinite} traces as one could also 
have done in the case of words.  
 If $D=\Ga \times \Ga$ the prefix metric on infinite real traces over $(\Ga, D)$ 
coincide with the preceding definition in the case 
of infinite words over $\Ga$.  

\hs If  $(\Ga, D)$ is a  dependence alphabet, a letter $a\in \Ga$ is said to be an isolated 
letter if $a$ is independent from all other 
letters of $\Ga$, i.e. 
 $\fa b\in \Ga-\{a\}$, $(a,b)\in I_D$.
 \nl From now on we suppose that a dependence alphabet has no isolated letter. Then the set 
$\mathbb{R}^\om(\Ga, D)$ of infinite real traces over  $(\Ga, D)$, equipped with the topology 
induced by the prefix metric, is homeomorphic to the Cantor set $\{0, 1\}^\om$, hence also 
 to  $\Sio$ for every 
finite alphabet $\Si$ having at least two letters \cite{kk}.  

\hs  Borel subsets of the Cantor set (hence also of topological 
spaces $\Sio$ or $\mathbb{R}^\om(\Ga, D)$)  form a strict infinite hierarchy, 
 the Borel hierarchy, which is defined from open sets by successive operations of countable 
unions and of countable intersections. We give the definition in the case 
of a topological space
$\Sio$, the definition being similar in the case 
of the topological space $\mathbb{R}^\om(\Ga, D)$. 
Then we recall some well known properties of Borel sets. 

\begin{Deff}
The classes ${\bf \Si_n^0}$ and ${\bf \Pi_n^0 }$ of the Borel Hierarchy
 on the topological space $\Sio$  are defined as follows:
\nl ${\bf \Si^0_1 }$ is the class of open subsets of $\Sio$.
\nl ${\bf \Pi^0_1 }$ is the class of closed subsets, 
i.e. complements of open subsets, of $\Sio$.
\nl And for any integer $n\geq 1$:
\nl ${\bf \Si^0_{n+1} }$   is the class of countable unions 
of ${\bf \Pi^0_n }$-subsets of  $\Sio$.
\nl ${\bf \Pi^0_{n+1} }$ is the class of countable intersections of 
${\bf \Si^0_n}$-subsets of $\Sio$.
\nl The Borel Hierarchy is also defined for transfinite levels.
The classes ${\bf \Si^0_\alpha }$
 and ${\bf \Pi^0_\alpha }$, for a non-null countable ordinal $\alpha$, are defined in the
 following way:
\nl ${\bf \Si^0_\alpha }$ is the class of countable unions of subsets of $\Sio$ in 
$\cup_{\gamma <\alpha}{\bf \Pi^0_\gamma }$.
 \nl ${\bf \Pi^0_\alpha }$ is the class of countable intersections of subsets of $\Sio$ in 
$\cup_{\gamma <\alpha}{\bf \Si^0_\gamma }$.
\end{Deff}

\begin{The}
\noi  
\begin{enumerate}
\ite[(a)] ${\bf \Si^0_\alpha }\cup {\bf \Pi^0_\alpha } \subsetneq  
{\bf \Si^0_{\alpha +1}}\cap {\bf \Pi^0_{\alpha +1} }$, for each countable 
ordinal  $\alpha \geq 1$. 
\ite[(b)] $\cup_{\gamma <\alpha}{\bf \Si^0_\gamma }= \cup_{\gamma <\alpha}{\bf \Pi^0_\gamma }
\subsetneq {\bf \Si^0_\alpha }\cap {\bf \Pi^0_\alpha }$, for each countable limit ordinal 
$\alpha$. 
\ite[(c)] A set $W\subseteq \Sio$ is in the class ${\bf \Si^0_\alpha }$ iff its 
complement is in the class ${\bf \Pi^0_\alpha }$. 
\ite[(d)] ${\bf \Si^0_\alpha } - {\bf \Pi^0_\alpha } \neq \emptyset $ and 
${\bf \Pi^0_\alpha } - {\bf \Si^0_\alpha } \neq \emptyset $ hold 
 for every countable  ordinal $\alpha\geq 1$. 
\end{enumerate}
\end{The}

\noi  We shall say that a subset of $\Sio$ is a Borel set of rank $\alpha$, for 
a countable ordinal $\alpha$,  iff 
it is in ${\bf \Si^0_{\alpha}}\cup {\bf \Pi^0_{\alpha}}$ but not in 
$\bigcup_{\gamma <\alpha}({\bf \Si^0_\gamma }\cup {\bf \Pi^0_\gamma})$. 

\hs Let us recall the  characterization of ${\bf \Pi^0_2 }$-subsets of $\Sio$, involving 
the $\delta$-limit $W^\delta$ of a finitary language $W$. 
For $W\subseteq \Si^\star$ and $\sigma\in \Sio$, 
  $\sigma \in W^\delta$ iff $\sigma$ has infinitely many prefixes in $W$, i.e. 
$W^\delta=\{\sigma\in \Sio / \exists^\om i$ such that $\sigma[i]\in W\}$ , see \cite{sta}.

\begin{Pro}
A subset $L$ of $\Sio$ is a ${\bf \Pi^0_2 }$-subset of $\Sio $ iff there exists 
a set $W\subseteq \Si^\star$ such that $L=W^\delta$.
\end{Pro}

\begin{Exa}\label{exa} Let $\Si=\{0, 1\}$ and $\mathcal{A}=(0^\star.1)^\om \subseteq \Sio$   
be  the set of 
$\om$-words over the alphabet $\Si$ with infinitely many occurrences of the letter $1$. 
It is well known that $\mathcal{A}$ is a ${\bf \Pi^0_2 }$-subset of $\Sio$ because 
$\mathcal{A}=((0^\star.1)^+)^\delta$ holds.
\end{Exa} 

\noi 
There are some subsets of the Cantor set, (hence also of the topological 
spaces $\Sio$ or $\mathbb{R}^\om(\Ga, D)$)   which are not Borel sets. There 
exists another hierarchy beyond the Borel hierarchy,  called the 
projective hierarchy. Projective sets  are defined  from  Borel sets by 
successive  operations of projection and complementation. 
We shall only need in this paper the first class of the projective hierarchy: the class 
 ${\bf \Si^1_1}$ of {\bf analytic} sets. 
A set $A \subseteq \Sio$ is analytic 
iff there exists a Borel set $B \subseteq (\Si \times Y)^\om$, with $Y$  a finite alphabet,  
such that $ x \in A \lra \exists y \in Y^\om $ such that $(x, y) \in B$, 
where $(x, y)\in (\Si \times Y)^\om$ is defined by:  
$(x, y)(i)=(x(i),y(i))$ for all integers $i\geq 1$.
\nl   Analytic sets  are also characterized  as continuous images of the Baire space $\om^\om$, 
which is the set of infinite sequences of non negative integers. It may be seen as the set 
of infinite words over the infinite alphabet $\om=\{0, 1, 2, \ldots \}$. The topology 
of the Baire space is then defined by a prefix metric which is just 
an extension of the previous one to the case of an infinite alphabet. 
\nl A set $A \subseteq \Sio$ (respectively $A \subseteq \mathbb{R}^\om(\Ga, D)$) 
is then analytic 
iff there exists a continuous function $f: \om^\om \ra \Sio$ 
(respectively $f: \om^\om \ra \mathbb{R}^\om(\Ga, D)$)
such that $f(\om^\om)=A$. 

\hs A ${\bf \Si^0_\alpha}$
 (respectively ${\bf \Pi^0_\alpha}$, ${\bf \Si^1_1}$)-complete set is a ${\bf \Si^0_\alpha}$
 (respectively ${\bf \Pi^0_\alpha}$, ${\bf \Si^1_1}$)- set which is 
in some sense a set of the highest 
topological complexity among the ${\bf \Si^0_\alpha}$
 (respectively ${\bf \Pi^0_\alpha}$, ${\bf \Si^1_1}$)- sets. 
 This notion is defined via reductions by continuous functions. 
More precisely a set $F\subseteq \Si^\om$ is said to be 
a ${\bf \Si^0_\alpha}$  (respectively ${\bf \Pi^0_\alpha}$, ${\bf \Si^1_1}$)-complete set 
iff for any set $E\subseteq Y^\om$  (with $Y$ a finite alphabet): 
 $E\in {\bf \Si^0_\alpha}$ (respectively $E\in {\bf \Pi^0_\alpha}$, ${\bf \Si^1_1}$) 
iff there exists a continuous 
function $f$ such that $E = f^{-1}(F)$.    ${\bf \Si^0_n}$ 
   (respectively  ${\bf \Pi^0_n}$)-complete sets, with $n$ an integer $\geq 1$, 
 are thoroughly characterized in \cite{stac}.  

\hs  
The \orl~ $\mathcal{A}=(0^\star.1)^\om$ given in Example \ref{exa} 
is a well known example of 
${\bf \Pi^0_2 }$-complete set.

\section{Rational languages of infinite traces}

\hs We want now to investigate the topological complexity of 
rational languages of infinite real traces. In a first step we shall give an upper 
bound of this complexity, showing that all rational languages 
$T \subseteq \mathbb{R}^\om(\Ga, D)$ are analytic sets. 
\nl We would like to use the characterization of rational languages 
$T \subseteq \mathbb{R}^\om(\Ga, D)$ given in item 3 of Theorem \ref{rat}:  
 $T=\varphi(L)$ for some \orl~ $L \subseteq \Gao$. Indeed every \orl~ is a Borel set 
(of rank at most 3) and the continuous image of a Borel set is an analytic set. 
Unfortunately,  the 
mapping $\varphi$ is not continuous as the following example shows. Let 
$(a, b)\in I_D$ and $x_n\in \Gao$ defined by $x_n=a^nba^\om$ for each integer $n\geq 1$. 
Then in $\Gao$ the sequence $(x_n)_{n\geq 1}$ is convergent and its limit is $a^\om$. 
But the sequence $(\varphi(x_n))_{n\geq 1}$ is constant in $\mathbb{R}^\om(\Ga, D)$ because 
for all $n\geq 1$ $\varphi(x_n)=\varphi(ba^\om)$. Thus the sequence $(\varphi(x_n))_{n\geq 1}$ 
is convergent but its limit is $\varphi(ba^\om)$ which is different from $\varphi(a^\om)$.  

\hs We shall use the characterization of rational languages 
$T \subseteq \mathbb{R}^\om(\Ga, D)$ given in item 2 of Theorem \ref{rat}:  
T is a finite union of sets of the form $R.S^\om$ where $S$ and $R$ are rational 
monoalphabetic languages of finite traces over $(\Ga, D)$ and $\varepsilon \notin S$. 
\nl We consider firstly such rational languages in the simple form $S^\om$ where $S$ is a  
monoalphabetic language of finite traces over $(\Ga, D)$ which does not contain the empty trace. 
The set $S$ is at most countable so it can be finite or countably infinite. 
In the first case card($S$)=$p$ and we can fix an enumeration of $S$ by a bijective function 
$\psi: \{0, 1, 2, \ldots , p-1\} \ra S$ 
and in the second case we can fix an enumeration of $S$ by a bijective function 
$\psi: \om=\{0, 1, 2, \ldots\} \ra S$.  
\nl Let now $H$ be the function defined from $\{0, 1, 2, \ldots , p-1\}^\om$ (in the first case) 
or from $\om^\om$ (in the second  case) into $\mathbb{R}^\om(\Ga, D)$ by:
$$H(n_1n_2\ldots n_i \ldots) = \psi(n_1).\psi(n_2)\ldots \psi(n_i)  \ldots$$
\noi for all sequences $n_1n_2\ldots n_i \ldots$ in $\{0, 1, 2, \ldots , p-1\}^\om$ 
(in the first case) or in  $\om^\om$ (in the second  case). It holds that 
$H(\{0, 1, 2, \ldots , p-1\}^\om)=S^\om$ (in the first case) or that $H(\om^\om)=S^\om$ 
(in the second  case). 
\nl It is easy to see that $H$ is a continuous function. A crucial point is that $S$ is a 
{\bf monoalphabetic} language, i.e. there exists $\Ga'\subseteq \Ga$ such that 
for all $s\in S$, $alph(s)=\Ga'$.  
 Let $\mathcal{N}=(n_i)_{i\geq 1}$ and 
$\mathcal{M}=(m_i)_{i\geq 1}$ be 
two infinite sequences of integers in $\{0, 1, 2, \ldots , p-1\}^\om$ or in  
$\om^\om$ such that for all  $i\leq k$ ~~ $n_i=m_i$. 
Then $ r \sqsubseteq H(\mathcal{N}) \lra r \sqsubseteq H(\mathcal{M})$ holds (at least) for all 
$r\in \mathbb{M}(\Ga, D) \mbox{ with } |r| \leq k$. 
Thus $l_{pref(H(\mathcal{N}) , H(\mathcal{M}) )} \geq k$
 and  $d_{pref}( H(\mathcal{N}) ,  H(\mathcal{M})  ) 
= 2^{-l_{pref( H(\mathcal{N})  ,  H(\mathcal{M})  )}} \leq 2^{-k}$. 
This implies that the function $H$ is continuous (and even uniformly continuous). 

\hs  If $S$ is finite,  then the set $S^\om$ is  the continuous image of the compact set 
$\{0, 1, 2, \ldots , p-1\}^\om$ thus it is a closed hence also an analytic 
subset of $\mathbb{R}^\om(\Ga, D)$. 
\nl If $S$ is infinite,  then the set $S^\om$ is  the continuous image of the Baire 
space $\om^\om$ thus it is an analytic set. 

\hs Let now $R \subseteq  \mathbb{M}(\Ga, D) $ be a language of finite traces. For $r\in R$ 
let $\theta_r: \mathbb{R}^\om(\Ga, D) \ra \mathbb{R}^\om(\Ga, D)$ be the function defined 
by $\theta_r(t)=r.t$ for all $t\in \mathbb{R}^\om(\Ga, D)$. 
It is easy to see that this function is continuous. Then $r.S^\om=\theta_r(S^\om)$ 
is an analytic set 
because the image of an analytic set by a continuous function is still an analytic set. 
The language $R.S^\om=\bigcup_{r\in R}r.S^\om$ is  a countable union of analytic sets 
(because $R$ is countable) but 
the class of analytic subsets of $\mathbb{R}^\om(\Ga, D)$ is closed under countable unions thus 
$R.S^\om$ is an analytic set. 
\nl  A rational language 
$T \subseteq \mathbb{R}^\om(\Ga, D)$ 
is a finite union of sets of the form $R.S^\om$ where $S$ and $R$ are rational 
monoalphabetic languages of finite traces over $(\Ga, D)$ and $\varepsilon  \notin S$. 
Then by finite union this language is an analytic set. 

\hs Notice that we have not used the fact that $S$ and $R$ are rational so the above proof 
can be applied to finite unions of sets of the form $R.S^\om$ where $S$ is a
monoalphabetic language of finite traces and we have got the following result.  

\begin{Pro}\label{ratan}
Let $(\Ga, D)$ be a dependence alphabet without isolated letter, and let $S_i$, $R_i$, 
$1\leq i\leq n$,  be  languages of finite traces over $(\Ga, D)$,  where, for all $i$,
  $S_i$ does not contain the empty trace and is 
monoalphabetic.  Then the language of infinite traces 
$$T = \bigcup_{1\leq i\leq n} R_i.S_i^\om$$ 
\noi is an analytic set. In particular every rational language 
$T \subseteq \mathbb{R}^\om(\Ga, D)$ is an analytic set. 

\end{Pro}

\noi  In order to prove the existence of 
${\bf \Si^1_1}$-complete rational language of infinite traces, 
 we shall use  results about languages of infinite binary trees whose nodes
are labelled in a finite alphabet $\Si$ having at least two letters.
\nl A node of an infinite binary tree is represented by a finite  word over 
the alphabet $\{l, r\}$ where $r$ means ``right" and $l$ means ``left". Then an 
infinite binary tree whose nodes are labelled  in $\Si$ may be viewed as a function
$t: \{l, r\}^\star \ra \Si$. The set of  infinite binary trees labelled in $\Si$ will be 
denoted $T_\Si^\om$.

\hs  There is a natural topology on this set $T_\Si^\om$  
which is defined by the following distance,  \cite{lt}. 
Let $t$ and $s$ be two distinct infinite trees in $T_\Si^\om$. 
Then the distance between $t$ and $s$ is $\frac{1}{2^n}$ where $n$ is the smallest integer 
such that $t(x)\neq s(x)$ for some word $x\in \{l, r\}^\star$ of length $n$.
\nl The open sets are then in the form $T_0.T_\Si^\om$ where $T_0$ is a set of finite labelled
trees. $T_0.T_\Si^\om$ is the set of infinite binary trees 
which extend some finite labelled binary tree $t_0\in T_0$, $t_0$ is here a sort of prefix, 
an ``initial subtree"
of a tree in $t_0.T_\Si^\om$.
\nl It is well known that the topological space $T_\Si^\om$ is homeomorphic to the Cantor 
set hence also to the topological spaces $\Si^\om$ or $\mathbb{R}^\om(\Ga, D)$. 

\hs The Borel hierarchy and the projective hierarchy on $T_\Si^\om$ are defined from open 
sets as in the cases of the topological spaces $\Si^\om$ or $\mathbb{R}^\om(\Ga, D)$.

\hs Let $t$ be a tree. A branch $B$ of $t$ is a subset of the set of nodes of $t$ which 
is linearly ordered by the tree partial order $\sqsubseteq$
 and which is closed under prefix relation, 
i.e. if  $x$ and $y$ are nodes of $t$ such that $y\in B$ and $x \sqsubseteq y$ then $x\in B$.
\nl A branch $B$ of a tree is said to be maximal iff there is no  other branch of $t$ 
which strictly contains $B$.

\hs Let $t$ be an infinite binary tree in $T_\Si^\om$. If $B$ is a maximal branch of $t$,
then this branch is infinite. Let $(u_i)_{i\geq 0}$ be the enumeration of the nodes in $B$
which is strictly increasing for the prefix order. 
\nl  The infinite sequence of labels of the nodes of  such a maximal 
branch $B$, i.e. $t(u_0)t(u_1)....t(u_n).....$  is called a path. It is an $\om$-word 
over the alphabet $\Si$.

\hs For $L\subseteq \Si^\om$ we denote $Path(L)$  the set of 
infinite trees $t$ in $T_\Si^\om$ such that $t$ has at least one path in $L$.

\hs It is well known that if $L\subseteq \Si^\om$ is a 
${\bf \Pi^0_2 }$-complete subset of $\Si^\om$ (or a Borel 
set of higher complexity in the Borel 
hierarchy) then the set $Path(L)$  is a ${\bf \Si^1_1 }$-complete subset of $T_\Si^\om$,  
\cite{niw85,simcras}, \cite[exercise]{pp}.

\hs In order to use this result
 we shall firstly code trees labelled in $\Si$ by  infinite words 
over the finite alphabet $\Ga = \Si \cup \Si' \cup\{A, B\}$ 
 where $\Si'=\{a' \mid a\in \Si\}$ is a disjoint copy of the alphabet $\Si$ and 
$A, B$ are  additional letters not in $\Si \cup \Si'$. 

\hs Consider now the set $\{l, r\}^\star$ of nodes of binary infinite trees.
For each integer $n\geq 0$, call $C_n$ the set of words of length $n$ of $\{l, r\}^\star$. 
Then $C_0=\{\varepsilon\}$, $C_1=\{l, r\}$, $C_2=\{ll, lr, rl, rr\}$ and so on.
$C_n$ is the set of nodes which appear at the $(n+1)^{th}$ level of an infinite binary tree.
The number of nodes of $C_n$ is $card(C_n)=2^n$.  We consider now  
the lexicographic order on $C_n$ (assuming that $l$ is before $r$ for this order).
Then, in the enumeration of the nodes with regard to this order, the  nodes of $C_1$ will 
be: $l, r$; the nodes of $C_3$ will be: $ lll, llr, lrl, lrr, rll, rlr, rrl, rrr$.
\nl Let $u^n_1,\ldots , u^n_j,\ldots  , u^n_{2^n}$ be such an enumeration of $C_n$ in 
the lexicographic order and let $v^n_1,..., v^n_j,..., v^n_{2^n}$ be the enumeration of the 
elements
of $C_n$ in the reverse order. Then for all integers $n\geq 0$ and $i$, $1\leq i\leq 2^n$, 
it holds that $v_i^n=u^n_{2^n+1-i}$.

\hs For  $t \in T_\Si^\om$ let $U_n^t=t(u^n_1)t(u^n_2) \ldots t(u^n_{2^n})$ be the finite word 
enumerating the labels of nodes in  $C_n$ in the lexicographic order, and let 
$V_n^t=t(v^n_1)t(v^n_2) \ldots t(v^n_{2^n})$ be the reverse sequence. 
Let $V_n^{'t} = \psi(V_n^t)$ where $\psi$ is the morphism from $\Si^\star$ into $\Si^{'\star}$ 
defined by $\psi(a)=a'$ for all $a \in\Si$. The code $g(t)$ of $t$ is then 

$$g(t) = V_0^{'t}.A.U_1^t.B.V_2^{'t}.A.U_3^t.B.V_4^{'t}.A \ldots 
 A.U_{2n+1}^t.B.V_{2n+2}^{'t}.A \ldots$$ 

\noi The $\om$-word $g(t)$  enumerates the  labels of the nodes 
of the tree $t$ which appear at successive levels $1, 2, 3,  \ldots$
The (images by $\psi$ of) labels of nodes occuring at odd level $2n+1$ are enumerated in the 
 reverse lexicographic order by the sequence $V_{2n}^{'t}$ and  
the labels of nodes occuring at even level $2n$ are enumerated in the 
  lexicographic order by the sequence $U_{2n-1}^{t}$. Labels of nodes of distinct levels are 
alternatively separated by a letter $A$ or a letter $B$.  

\hs Let now $(\Ga, D)$ be a dependence alphabet where $\Ga = \Si \cup \Si' \cup\{A, B\}$ 
 and the independence relation  
$I_D = \Ga \times \Ga - D$ is defined by 

$$I_D= \Si \times (\{A\}\cup \Si') ~ \bigcup ~(\{A\}\cup \Si') \times \Si ~ \bigcup ~
\Si' \times \{B\} ~ \bigcup ~ \{B\} \times \Si'$$

\noi i.e. letters of $\Si$  may only commute with $A$ and with letters in $\Si'$ while letters 
 of $\Si'$  may only commute with $B$ and with letters in $\Si$, letter $A$ (respectively 
$B$) may only commute with letters in $\Si$ (respectively, with letters in $\Si'$).

\hs Let now $h: T_\Si^\om   \ra  \mathbb{R}^\om(\Ga, D)$ be 
the function defined by: 
$$\fa t \in T_\Si^\om ~~~~ h(t)=\varphi(g(t))$$ 
\noi  We firstly state the following result. 

\begin{Lem}
The above defined function $h: T_\Si^\om   \ra  \mathbb{R}^\om(\Ga, D)$ is continuous.
\end{Lem}

\proo  Let us remark that in a segment  
$$B.V_{2n}^{'t}.A.U_{2n+1}^t.B.V_{2n+2}^{'t}.A$$
\noi of an    $\om$-word $g(t)$ written as above,  letters of $U_{2n+1}^t$ may only 
commute with the preceding letter $A$ and letters of $V_{2n}^{'t}$. In a similar manner 
 letters of $V_{2n+2}^{'t}$ may only 
commute with the preceding letter $B$ and letters of $U_{2n+1}^t$. 

\hs So if two infinite binary trees $t, s \in T_\Si^\om$ have the same labels on their 
$k$ first levels, ($k>1$), then for all  $r\in \mathbb{M}(\Ga, D)$ such that 
$$|r| \leq (k-1) + 1+2+2^2+ \ldots +2^{k-2}$$ 
\noi it holds that 
$$r \sqsubseteq h(t) \lra r \sqsubseteq h(s)$$ 
\noi So 
$$l_{pref(h(t), h(s))} \geq  (k-1) + 1+2+2^2+ \ldots +2^{k-2} \geq   2^{k-1}$$
\noi and 
$$d_{pref}((h(t), h(s))=2^{-l_{pref(h(t), h(s))}} \leq 2^{-2^{k-1}}$$ 

\hs so we have proved:
$$ \fa t, s \in T_\Si^\om ~~~~~ d(t, s)\leq 2^{-k} \ra  
d_{pref}((h(t), h(s)) \leq 2^{-2^{k-1}}$$
\noi  Thus the function $h$ is continuous  (and 
even uniformly continuous). \ep

\hs Let now  $\mathcal{R} \subseteq \Sio$ be a regular \ol . 
We are going to define from  $\mathcal{R}$
a  language of infinite real traces
$\mathcal{L}$  over the  dependence alphabet $(\Ga, D)$ defined above. Then we shall prove 
that  $\mathcal{L}$  is rational and 
 that  $Path(\mathcal{R}) = h^{-1} (\mathcal{L})$.  

\hs Let us firstly define  $\mathcal{L}$ as being 
 the set of infinite traces $\varphi(\sigma)$ where 
$\sigma\in \Gao$ may be written 
in the following form:

$$\sigma=x(1).u_1.A.v_1.x(2).u_2.B.v_2.x(3).u_3.A \ldots 
 A.v_{2n+1}.x(2n+2).u_{2n+2}.B.v_{2n+2}.x(2n+3).u_{2n+3}.A\ldots $$ 

\hs where for all integers $i\geq 0$, 
$$x(2i+2)\in \Si  ~~~~ \mbox{  and }  ~~~~ x(2i+1)\in \Si'$$ 
$$u_{2i+2}, v_{2i+1} \in \Sis ~~~~ \mbox{  and }  ~~~~ u_{2i+1}, v_{2i+2} \in \Si^{'\star}$$
$$|v_i|=2|u_i|  ~~~~ \mbox{  or }  ~~~~|v_i|=2|u_i|+1$$ 
and the $\om$-word 
\nl $x=\psi^{-1}(x(1))x(2)\psi^{-1}(x(3)) \ldots x(2n)\psi^{-1}(x(2n+1))x(2n+2)\ldots $  
is in $\mathcal{R}$.  

\begin{Lem}\label{lrat}
The above defined language $\mathcal{L}$ of infinite real traces is rational.  
\end{Lem}

\proo Every $\om$-word 

$$\sigma=x(1).u_1.A.v_1.x(2).u_2.B.v_2.x(3).u_3.A \ldots 
 A.v_{2n+1}.x(2n+2).u_{2n+2}.B.v_{2n+2}.x(2n+3).u_{2n+3}.A\ldots $$ 
\noi written as above is equivalent, 
modulo the equivalence relation $\sim_{I_D}$ over infinite words in $\Gao$, to the 
infinite word 
$$\sigma'=x(1).u_1.v_1.A.x(2).u_2.v_2.B.x(3).u_3.v_3.A \ldots 
 A.x(2n+2).u_{2n+2}.v_{2n+2}.B.x(2n+3).u_{2n+3}.v_{2n+3}.A\ldots $$ 
\noi But letters of $\Si$ may commute also with letters of $\Si'$ and for all 
integers $i$, 
$$|v_i|=2|u_i|  ~~~~ \mbox{  or }  ~~~~|v_i|=2|u_i|+1$$ 
\noi  by definition of $\mathcal{L}$. Thus every $\om$-word $\sigma$ written as above is also 
equivalent, modulo $\sim_{I_D}$, to an infinite word in $\Gao$ in the form

$$\sigma''=x(1).W_1.A.x(2).W_2.B.x(3).W_3.A \ldots 
 A.x(2n+2).W_{2n+2}.B.x(2n+3).W_{2n+3}.A\ldots $$ 
\noi where for all integers $i\geq 0$,
$$W_{2i+1} \in (\Si'\Si^2)^\star.(\Si\cup\{\varepsilon\})$$
$$W_{2i+2} \in (\Si\Si^{'2})^\star.(\Si'\cup\{\varepsilon\})$$

\noi Let  $L$ be the $\om$-language over the alphabet $\Ga$ formed by all such  
$\om$-words $\sigma''$ such that
$x=\psi^{-1}(x(1))x(2)\psi^{-1}(x(3)) \ldots x(2n)\psi^{-1}(x(2n+1))x(2n+2)\ldots $  
is in $\mathcal{R}$. 
\nl It is easy to see that $L$ is an \orl. Moreover  $\mathcal{L}=\varphi(L)$ thus 
we can infer from Theorem \ref{rat}  that $\mathcal{L}$ is a rational language of infinite real 
traces. 

\ep 

\hs We are going now to prove the following result. 

\begin{Lem}\label{lem} For $\mathcal{L}$ defined as above from the \ol~ $\mathcal{R}$,  it holds that  
 $Path(\mathcal{R}) = h^{-1} (\mathcal{L})$, i.e. 
~~~~
$\fa t \in T_\Si^\om ~~~~~  h(t) \in \mathcal{L} \longleftrightarrow t \in Path(\mathcal{R})$.  
\end{Lem}

\proo Suppose that $h(t) \in \mathcal{L}$ for some $t \in T_\Si^\om$. Then 
$h(t)=\varphi(g(t))=\varphi(\sigma)$ where 
$\sigma\in \Gao$ may be written 
in the following form:

$$\sigma=x(1).u_1.A.v_1.x(2).u_2.B.v_2.x(3).u_3.A \ldots 
 A.v_{2n+1}.x(2n+2).u_{2n+2}.B.v_{2n+2}.x(2n+3).u_{2n+3}.A\ldots $$ 

\hs where for all integers $i\geq 0$, 
$$x(2i+2)\in \Si  ~~~~ \mbox{  and }  ~~~~ x(2i+1)\in \Si'$$ 
$$u_{2i+2}, v_{2i+1} \in \Sis ~~~~ \mbox{  and }  ~~~~ u_{2i+1}, v_{2i+2} \in \Si^{'\star}$$
$$|v_i|=2|u_i|  ~~~~ \mbox{  or }  ~~~~|v_i|=2|u_i|+1$$ 
and the $\om$-word 
\nl $x=\psi^{-1}(x(1))x(2)\psi^{-1}(x(3)) \ldots x(2n)\psi^{-1}(x(2n+1))x(2n+2)\ldots $  
is in $\mathcal{R}$.  

\hs Then it is  easy to see that $\sigma=g(t)$, because of the definition of $\sigma$, 
of $g(t)$, and of the independence relation $I_D$ on $\Ga$. 
\nl  Then $\psi^{-1}(x(1))=t(v_1^0)$ and $u_1=\varepsilon$, 
then $|v_1|=2|u_1|=0$ or  $|v_1|=2|u_1|+1=1$. 
If  $|v_1|=0$ then $x(2)=t(u_1^1)$ and if $|v_1|=1$ then $x(2)=t(u_2^1)$. Then the choice of 
 $|v_1|=2|u_1|$ or of $|v_1|=2|u_1|+1$ implies that $x(2)$ is the label 
of the left or the rigth successor of the root node $v_1^0=\varepsilon$. 
\nl  This phenomenon will happen for next levels. The choice of 
$|v_i|=2|u_i|$ or of $|v_i|=2|u_i|+1$ determines one of the two successor nodes 
of a node at level $i$ 
(whose label is $x(i)$ if $i$ is even or $\psi^{-1}(x(i))$ if $i$ is odd) and then 
the label of this successor is  $\psi^{-1}(x(i+1))$ if $i$ is even, or $x(i+1)$ if $i$ is odd. 
\nl Thus the successive choices determine a branch of $t$   
and the labels of nodes of this branch (changing only $x(2n+1)$ in $\psi^{-1}(x(2n+1))$) 
  form a path 
$x=\psi^{-1}(x(1))x(2)\psi^{-1}(x(3)) \ldots x(2n)\psi^{-1}(x(2n+1))x(2n+2)\ldots $  
which is in $\mathcal{R}$. Then $t\in  Path(\mathcal{R})$. 

\hs Conversely it is easy to see that if $t\in  Path(\mathcal{R})$, the infinite word 
$g(t)$ may be written as a word $\sigma$  in the above form. Then 
$h(t)=\varphi(g(t))=\varphi(\sigma)$ is in $\mathcal{L}$. 
\ep 

\hs  We can now state the following 

\begin{The} There exist some ${\bf \Si^1_1}$-complete, hence non Borel, 
 rational languages of infinite real traces. 
\end{The}

\proo  Suppose  $\mathcal{R}\subseteq \Sio$  is   a ${\bf \Pi_2^0}$-complete  \orl . 
Let then $\mathcal{L}\subseteq \mathbb{R}^\om(\Ga, D)$ be defined as above. 
$\mathcal{L}$ is a rational language of infinite real traces by Lemma \ref{lrat}. 
Then $\mathcal{L}$ is an analytic subset of $\mathbb{R}^\om(\Ga, D)$ 
by Proposition \ref{ratan}. 
\nl But $Path(\mathcal{R})$ is a  ${\bf \Si^1_1}$-complete set and 
$Path(\mathcal{R}) = h^{-1} (\mathcal{L})$ holds by Lemma \ref{lem} thus  
$\mathcal{L}$ is also ${\bf \Si^1_1}$-complete. 
In particular  $\mathcal{L}$ is not a Borel set.  \ep 

\section{Concluding remarks}

\noi The existence of a ${\bf \Si^1_1}$-complete infinitary rational relation has been used 
 to get many undecidability results in \cite{rel-dec} and the existence 
of a ${\bf \Si^1_1}$-complete 
context free $\om$-language led to other undecidability results in \cite{finb,ambcf,fs}. 
In particular,  the topological complexity and the degree of ambiguity of an 
infinitary rational relation or of a context free $\om$-language are 
highly undecidable. 
\nl In a similar way,  the existence of a ${\bf \Si^1_1}$-complete  recognizable language 
of infinite pictures, proved in \cite{atw} by  Altenbernd, Thomas, and 
Wöhrle, has been used in \cite{rec-pic} to 
prove many undecidability results, giving  in particular the answer to some open questions 
of \cite{atw}. 
\nl Topological arguments following from the existence of ${\bf \Si^1_1}$-complete rational 
languages of infinite real traces
can also be used to prove similar undecidability results for languages of infinite traces. 
\nl In \cite{fs}  have been established some links between the existence of 
a ${\bf \Si^1_1}$-complete $\om$-language in the form $V^\om$ and the number of decompositions 
of $\om$-words of $V^\om$ in words of $V$. 
\nl We think that such facts could be useful in the domain of combinatorics of traces. 
The code problem for traces is important in Trace Theory and several questions are 
still opened \cite{hm}. 
The analogue of the notion of $\om$-code and the study of the number of decompositions 
of infinite traces of $V^\om$, where $V$ is a set of  finite traces, in infinite 
product of traces of $V$, is also an important subject related to practical applications 
and to the notion of ambiguity (see \cite{aug1,aa} for related results in the case of words). 
We think that topological arguments could be useful in this 
research area, and the existence of several ${\bf \Si^1_1}$-complete languages 
of infinite traces, having different structures, could be useful in the cases of different 
dependence alphabets.

\hs {\bf  Acknowledgements.}  Thanks to  the anonymous referees 
for useful comments on a preliminary version of this paper.

\end{document}